# Towards Applying Deep Learning to The Internet of Things: A Model and A Framework

Samaa Elnagar [1,*], Kweku-Muata Osei-Bryson [1]

[1] Information Systems, Virginia Commonwealth University, Richmond, USA

{elnagarsa,kmosei}@vcu.com

**Abstract.** Deep Learning (DL) modeling has been a recent topic of interest. With the accelerating need to embed Deep Learning Networks (DLNs) to the Internet of Things (IoT) applications, many DL optimization techniques were developed to enable applying DL to IoTs. However, despite the plethora of DL optimization techniques, there is always a trade-off between accuracy, latency, and cost. Moreover, there are no specific criteria for selecting the best optimization model for a specific scenario. Therefore, this research aims at providing a DL optimization model that eases the selection and re-using DLNs on IoTs. In addition, the research presents an initial design for a DL optimization model management framework. This framework would help organizations choose the optimal DL optimization model that maximizes performance without sacrificing quality. The research would add to the IS design science knowledge as well as the industry by providing insights to many IT managers to apply DLNs to IoTs such as machines and robots.

## 1    Introduction

With the emergence of Artificial Intelligence (AI) especially Deep Learning (*DL*) methods, there has been much focus on giving IoT devices the knowledge and the inference capabilities of humans. Internet of Things (*IoT*s) could be in the form of *End* and *Edge* devices. a Deep Learning Networks (*DLNs*) could be applied to various *IoT*s such as robotics, self-driving vehicles, augmented reality, and digital assistance [1].

On the one hand, *DL* models are known for their computational cost and complexity, so they are mostly run on servers on the cloud [2]. On the other hand, *end devices (e.g. smart sensors)* and *edge devices* (e.g. routers) are often battery powered, have limited memory, processing and energy resources to store and process data. Applying *DL* models to *IoT* is challenging, in a manner that is similar to attempting to fit a giant elephant in a tight limited tunnel. DLNs have to be optimized and compressed to fit to IoTs limited computational sources. In addition, the optimization of the IoTs themselves is also necessary in terms of memory and hardware optimization [3].

*IoT*s provide businesses with the leverage of allowing data generated from customers to be included in their decision-making processes [4] that led to the development of the *Edge Computing* paradigm [5]. The development of optimization algorithms for *DLN*s could empower *IoT* with many cognitive abilities [6]. However, there is a lack of guidance on how to choose the appropriate *DLN* and optimization model to be applied according to different settings. Therefore, the main focus of this research is the development of a model for optimizing *DL* to be applied on *End* and *Edge* devices? Our two major objectives are:

1. To build an *DL* optimization model that provides an end-to-end modeling for applying *DLN* to IoTs according to specific contextual settings.

2. To provide the foundational bricks for building a Deep Learning Optimization Model Management Framework (DLOM)[2] that maximizes value gained from applying the presented model.

The research adds to IS body of knowledge by creating an optimization model that connects two emergent yet important paradigms: *DL* and *IoT*. In addition, the paper sheds light on the issues related with optimizing *DL* models for *IoT*. The target audience for this research are IT practitioners who are eager to apply the *DL* to edge and end devices, but they are concerned about how to choose the convenient models to be applied to their environments.

Why Is There A Need for DL Modeling for IoT? *DL* modeling is fueled by the increasing need to embed *DL* to *IoT* (end and edge devices) and the complexity of transferring *DL* to such limited capacity devices [7]. In addition, changing a DL model (its network structure and hyper-parameters) to fit a customized environment is an exhaustive empirical inquiry [8].

Applications of *DLN*s to *IoT* are endless. For example a small robot that applies object recognition could double the production rates and reduce human errors significantly [9]. In order to embed *DLN*s to *IoT*s (edge and end devices), optimizing *DLN*s is necessary. *DLN*s optimization targets not only reducing DL model size, but also optimizing the memory and computational requirements of *IoT*s to accelerate the inference on these limited devices.

## 2 Background & Review of Previous related Research

### 2.1 Overview on Edge computing and Internet of Things (IoT)

According to O'Connor [10], the *IoT* refers to "computing devices often with sensor capability to collect, share, and transfer data using the Internet" (p. 80). *IoT*s have many strategic benefits such as increasing automation and error rates, enhancing trust in asset management and providing greater predictability in risk-based decision-making [11]. The success of the *IoT*s and the huge amount of data generated from these devices created the need for *Edge computing* [5] in which data processing are performed near network edge, rather than centralized computing on the cloud. *Edge Computing (EC)* was developed to solve latency, network decency, costs, security, and privacy [12].

### 2.2 Overview on Deep Learning Model Management Systems (DL-MMS)

Research on model management in the area of edge computing and *IoT* are relatively few. A Model Management System (*MMS*) supports the creation, compilation, reuse, evolution, and execution of mappings between schemas represented in a wide range of meta-models [13]. The most challenging research question in developing *MMS* is how to support mappings between many popular metamodels.

Gupta [14] developed iFogSim to model *IoT* and Fog environments to measure the impact of resource management techniques in terms of latency, network congestion, energy consumption, and cost. However, the iFogSim didn't address the modeling of *DLNs* nor IoT as hardware. The research only focused on resource management policies in terms of network, RAM consumption, and execution time.

Ko et al.[15] presented a modeling framework for *Edge Computing (EC)* that provides useful guidelines, provisioning and planning. Unfortunately, the research focused on networking topologies only. There have been some recent attempts to address *DL* model management in both academia and industry, such as ModelHub , ModelDB , MLflow [8]. Unfortunately, most of these approaches either require a considerable amount of customization or they are limited to a specific commercial platform. In addition, none of them are targeting *IoT* devices.

ModelDB [16] is one of the early systems that aimed at addressing *DL* model management issues, and it comes very close to our solution in its functionality. However, ModelDB is tailored for specific machine learning models, and provides limited support for *DLNs*. ModelHub [8] is a high-profile deep learning management system that proposes a domain specific language to allow easy exploration of models, a model versioning system, and a deep-learning-specific storage system. It also provides a cloud-based repository. Schelter et al., [17] provided an automated tool to extract the model's metadata with an interactive visualization to query and compare experiments.

### 2.3 Gap in Literature

The findings of literature could be summarized in two directions: modeling for *EC* as a networking paradigm which moves computation near end devices; and modeling of the deep learning models to be applied to devices with high computational resources. However, none of the modeling techniques aggregated *DL* optimization for *IoT*. *DL* optimization modeling is more complicated than general *DL* modeling, where choosing the network best hyper parameter is just one sub problem in the *DL* optimization schema. Therefore, the proposed model management system is the first to provide modeling for *DL* optimization for *IoT*.

### 2.4 Objectives for DL Optimization

In this section, we are pointing out the pillars for *DL* Optimization which are *Privacy and Security, Compression, Quantization, and Hardware Optimization*

**Privacy and Security:** There is always a tradeoff between privacy, computational complexity and response delay as shown in Figure 1. On the server side, shielded execution or assigning a secure enclave for custom models to be trained on the cloud server [18]. On the edge side, CryptoNets [19] and fog-based nodes are used to encrypt data.

**Compression:** compression aims to reduce the massive size of *DL* networks. One of the popular compression methods is the *Pruning* technique that eliminates the connections between neurons to directly reduce the feature map width and shrink the network size. However, removing neurons might be dramatically challenging because it will change the input of the following layer [20]. *Tensor Decomposition* (*TD*) is also used to further reduce the network weight especially for convolutional kernels which can be viewed as a 4D tensors. *TD* is derived by the intuition that there is a significant amount of redundancy in the 4D tensor [21].

*Fine-tuning* is another method used to train custom models with a generative objective, followed by an additional training stage with a discriminative objective. The underlying assumption is that a reasonably good result on the large training data set already puts the network near a local optimum in the parameter space so that even a small amount of new data is able to quickly lead to an optimum [22]. *Knowledge Distillation (KD)* [23] is another compression technique that involves training a quantized neural network (student model) with the help of a full-precision pre-trained network (teacher model). The compressed student model can take the benefit of transferring knowledge from the teacher model.

**Quantization**: Quantization aims at compacting the number of bits required to store the *DLN* weights usually from 64 bit to 8 bits [24]. Quantization of the *DLN* without training is a fast process but the accuracy of the resultant network is particularly low compared to quantized networks after training [25] . However, it remains an open problem as to what is the best level of quantization that won't hurt accuracy for a given network [26].

**Hardware Optimization:** Since *IoT*s are resource limited, hardware specs of *IoT* must be chosen carefully. The adoption of Field Programmable Gate Arrays (FPGA) offers 2-bit ternary and 1-bit binary *DLN*s which resulted in as high as 90% by pruning because FPGAs designed for extreme customizability [3].

Google's Tensor processing Unit (*TPU*) is powering a wide range of Google real time services. *TPU* often delivers 15x to 30x faster inference than CPU or GPU, and even more per watt power at a comparable cost level. Its outstanding inference performance originates from major design optimizations: Int8 quantization, *DNN*-inference-specific CISC instruction set, massively parallel matrix processor, and minimal deterministic design [27].

### 2.5    Knowledge Management Challenges for DL Optimization

The knowledge acquired by *DL* is in the form of tacit knowledge that cannot be converted to a mathematical formula or a logical model to be applied to other different problems. Let's at first summarize what are the major objectives for optimizing *DL* models as mentioned in the overview discussed earlier. We can conclude that there are six main objectives. Three are to be maximized which are *performance, reliability and security*. Another three are to be minimized which are *cost, latency, and complexity* as shown in Figure1. Unfortunately, there is no single solution that could overcome all these challenges. A solution might have a positive effect on one objective but negative effect on others as shown in Table 1. For example: pruning as an optimization technique enhances performance but increases latency. So, there is always a tradeoff between different objectives (e.g tradeoff between performance and cost).

To elaborate the optimization methodologies and related issues, we divided the methodologies on the cloud side or the server side and the client side or the *IoT* side as shown in Figure 2. The green shapes are the optimization methodologies and the red circles are the issues related to these methodologies. So, in case of no optimization method is used, bottlenecks, latency and performance degradation will occur [28]. When fine tuning is applied, a *discriminative objective* function should be applied, otherwise we will have training imbalance [29]. Compression techniques are trying to decrease the *DLN* size by shrinking weights, connection and layers. However, removing too much of the network could affect the *DLN* throughput and response time. Shielding for the company *DLN*s on cloud servers provides security but also increases performance overhead.

On the client side, optimizing the IoT device could be achieved by choosing a powerful processor, decent memory, and fast communication channels. However, without correct optimization on the cloud side, there are potential memory allocation problems, inference latency which decrease throughput and traffic overhead [30].

**Table 1**: Effects of Optimization Methods on different DL Optimization Objectives.

| Optimization Method | Performance | Latency Reduction | Cost Reduction | Complexity Reduction | Reliability | Privacy |
|---|---|---|---|---|---|---|
| Pruning | + | - | + | + | + | 0 |
| Knowledge Distillation | - | - | + | + | + | 0 |
| Quantization | - | + | + | + | - | 0 |
| Fog Computing | + | - | - | - | + | + |
| Shielded Execution | + | - | - | - | + | + |
| Tensor Decomposition | - | + | + | + | - | 0 |
| Hardware Optimization | + | + | - | - | + | 0 |

**(+) means to increase; (-) means to decrease; (0) means has no effect.**

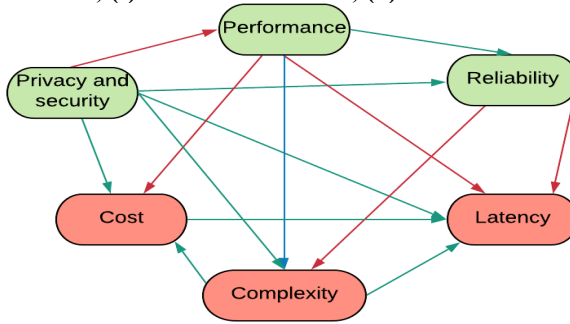

**Fig. 1**: DL Optimization Objectives.

## 3  Research Methodology

We are following a design science methodology called *complexity control* perspective for the design of the *DL* optimization Model [31]. This perspective is built upon the uncertainty of understanding the sociotechnical nature of complex systems. Therefore, we consider the proposed design as an initial iteration in a series of upcoming design iterations where the design is subject to change in each iteration. Based on the *complexity control* perspective, kernel theories are used to just predict how a particular design artifact will perform in the application environment. The supporting theoretical perspective for building the artifact should focus on the *information exchange* [32] and effective modeling.

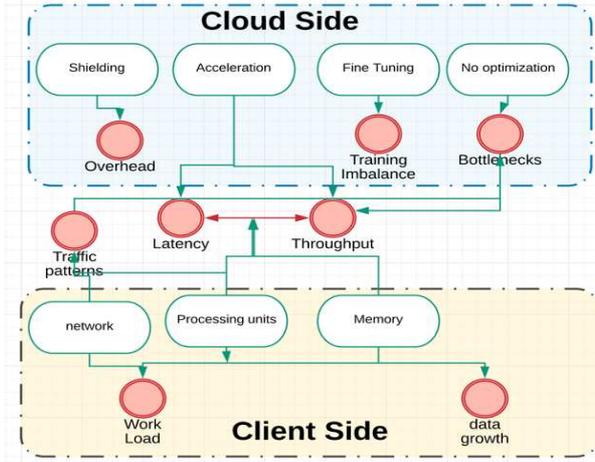

**Fig. 2**: Issues with each Optimization Methodology.

To establish the rigor of the research, we built the DL optimization modeling schemas based on the established model of DM³ ontology [13]. In addition, the developed schema not only contains the technical methodologies used in the optimization process, but also encompass quality metrics to evaluate different models [33]. The summary of the research activities of this research based on [34] design guidelines is summarized in Table 2:

**Table 2:** DSR Guidelines-based Activities of this Research

| DSR Guideline | Activity of this Research Project |
|---|---|
| Design as an Artefact | Development of a DL optimization modeling schema for modeling DLNs application to IoT. |
| Design Evaluation | in this stage of the research program, the artefacts will be evaluated using illustrative example |
| Research Rigor | Building the artefact based on established theories and utilization of established modeling techniques such as DM³ [13] |
| Design as a Search Process | Research on DL, IoT, MMS, EC and other relevant literature in order to identify appropriate techniques & other results that could be used to inform the design of the procedure |

## 4    Proposed Solution

### 4.1    The DL Optimization Model Schema

The *DL* optimization modeling lifecycle exposes several knowledge management challenges such as: a) managing many different models and their settings, b) large storage footprints of learned parameters, c) comparing models, d) selecting the best optimization techniques both in server and cloud side, and e) sharing models with others. Therefore, the first essential step is building a unified schema that structures the optimization process and eases the management processes. Abstracting the main classes of the *DL* optimization schema, it will be divided into modeling the *DLN* itself, modeling optimization techniques, and modeling end and edge device hardware as shown in Figure 3.

The proposed schema consists of six classes as shown in Table 3. However, the developed modeling schema is subject to change through the design iterations according to the *complexity control* perspective discussed earlier.

The first class of the *DL* optimization modeling schema is the *Model Class* that encompasses information about the previous models (meta-data) such as the date the model was built, the purpose (business focus), and the total cost. The *Model Class* is to ease the compare and contrast of different models and retrieval of models based on business focus and planned cost. The most important attribute in the *Model Class* is *Rating*. The *Rating* attribute is provided by the users who rate the model based on six objectives of *performance, reliability, security, cost, latency, and complexity*. All objectives have a fixed measure scale from 1 to 5 where 1 is the worst and 5 is the best. The first three objectives (i.e. *performance, reliability,* and *security*) are forward scaled, while the other three (i.e. *cost, latency, and complexity*) are reverse scaled. The rating attribute will help new users prioritize their preferences and it will ease the selection of the appropriate model.

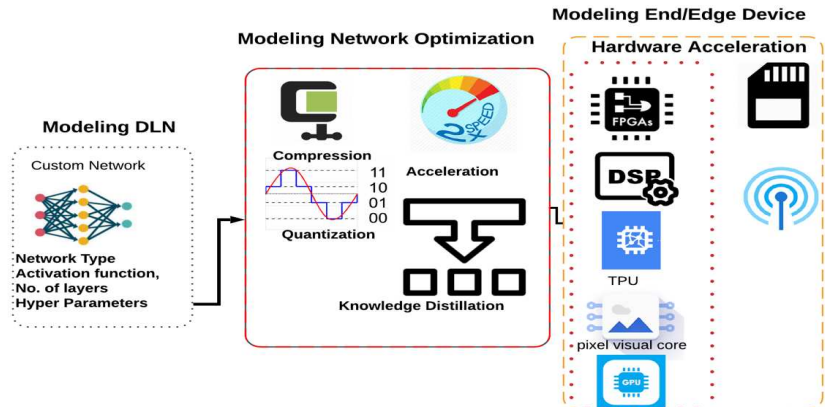

**Fig. 3:** DL Optimization Model Components.

The *Cloud Configuration Class* encompasses the settings for the main *DLN*s on the cloud such as security protocols and cost plan. This class is important as running models on the cloud is costly and prone to security threats. The *End devices Specifications Class* provides information on the *IoT* technical specifications. The *Main DLN Class* aims at modeling the DLN specifications on the cloud which includes the network name, no. of layers, weights, hyper parameters, activation, and loss functions. For example: *DLN* is ResNet-50, activation is SoftMax, input and output layer size are 300.

The *Optimization Class* records the optimization techniques used such as the Tensor Decomposition and Quantization. The *Optimization Class*, the *Cloud Configuration Class and Main DLN Class* target the IT specialists to gain information about the technical requirements needed to implement DL optimization models. The *DL* Optimization model schema could be visualized as an ontology or a knowledge graph as in Figure 4.

### 4.1.1 Performance Measures Class

The *Performance* class might be the most important class in the *DL* optimization model schema that nominates different models over others. The performance class is built upon the previous performance criteria discussed in the literature:

- The root-mean-square error (RMSE), mean absolute relative difference (MARD), and Mean Average Precision (mAP) that serve as the primary indicators to evaluate DLN accuracy [1].
- The vital performance goal of *DL* modeling for *IoT* is to decrease inference latency. Automated services relying on inference are required to respond in near real time. For example: self-driving cars require less than 200ms inference time [27]. Latency is divided into two metrics: inference time and system response time.
- Throughput: measures the completed work amount against the time consumed. It also used to measure the performance of a processor, memory or network interaction [18].
- Energy watts to be measured in Watts, memory is measured in MB.
- Stability [35](variance in accuracy for the same target). Stability could be measured by calculating the variance in the average accuracy measured every day for a certain period.

**Table 3:** Basic Classes of the DL Optimization Model Schema

| Performance Class | End devices Specs. Class | Main DLN Class | Optimization Class | Cloud Conf. Class | Model Class |
|---|---|---|---|---|---|
| System Latency | Name | Name | Quantization | Host address | Year created |
| Inference Latency | CPU, GPU | Training Dataset | KD | Response time | Rating |
| Accuracy | Memory | Hyper parameters | TD | Shielded execution | Application area |
| Stability (variance in accuracy for the same target) | Camera | Activation/ loss functions | Pruning | Security protocols | Total cost |
| Average power consumption | DL mobile Framework | No of layers | Fine-tuning | Cost plan | Purpose |
| Throughput | Price | No. of input / outputs | Algorithms | Backup-Address | No. IoT devices |

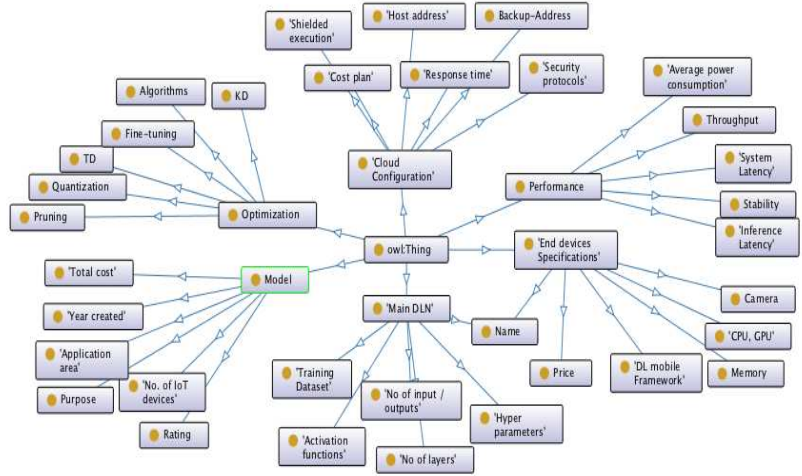

**Fig. 4:** the ontological representation DL Optimization Model Schema

# 5 The DL for IoT Model Management Framework (DLOM²)

Since *DL* optimization for *IoT* is still an emerging field, our model management framework might be one of the earliest attempts to address *DL* optimization for *IoT*. The model management system is not just a knowledge base for *DL* models, but also a prediction platform. Due to the limited knowledge about how to mix and match different optimization techniques to maximize *DL* modeling objectives, existing amalgamation of optimized *DL* techniques represents a rich knowledge repository that can be used to analyze, explore and create new models.

This knowledge repository keeps the optimized *DL* models saved according to the *DL* optimization schema presented earlier. The knowledge repository will ease querying of models based on different criteria using SPARQL. The architecture of DLOM² consists of four main components: *Cloud-based Repository, Graphical User Interface (GUI), Decision support system (DSS)*, and *a DL Modeling Network* as shown in Figure 5.

## 5.1 Graphical User Interface (*GUI*)

The *GUI* is responsible for providing an interactive user interface to help them select the best *DL* model. The *GUI* is responsible for taking modeling requests and displaying results to the user. In the modeling request, several criteria will be chosen to select the best optimization *DL* model. The modeling request is sent to the *DSS* to select the best model. The *GUI* is also responsible for displaying suggested *DL* models along with explanation for the suggestion made.

## 5.2 Cloud Repository

The cloud repository is responsible for storing different *DL* optimization models based on the *DL* optimization schema provided earlier. However, it might be very complicated with all the information in a single repository. For example: it is not wise to store *DL* hyper parameters (technical, hard to read knowledge) along with performance metrics (human friendly knowledge). Therefore, there are four different repositories where each repository is responsible for storing certain classes to ease *management, maintenance and replication*. The four repositories are the DLN parameters repository, the client-side configurations repository, the server-side configurations repository, and the model and performance repository.

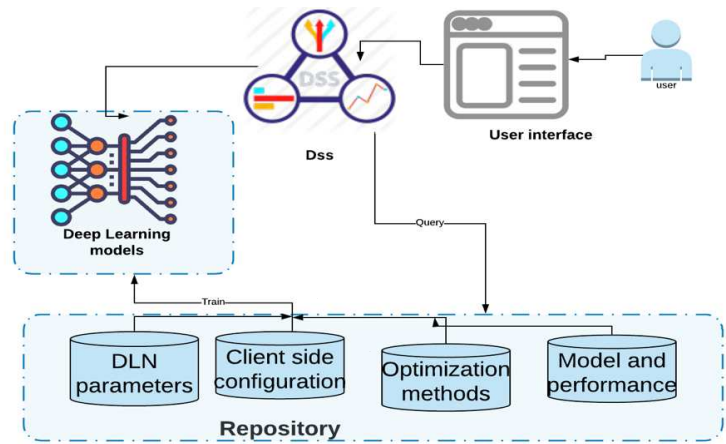

**Fig. 5:** The DLOM² Architecture.

### 5.3 The DL Modeling Network

Since there are no predefined criteria to design the best optimization *DL* models, an *unsupervised DL* approach is needed to explore the patterns of the development of different *DLN* optimization models. Training a *DL* model will help to understand how *DLN* optimization models could be created. So, a *DL* network is trained with different optimization models previously stored in the repository. The optimization models are saved in the repository according to the schema introduced in the previous section. An example of how the *End devices Specs* class will look like according to the schema:

*<**End_devices_Specs**> <Name>Raspberry pi 3</Name> <price>70</price> <DLFramework> MobileNet V3</DLFramework> <Memory>8 GB</Memory> <Camera>16 MP </Camera><CPU> </CPU> **</End_devices_Specs>***

This model will be the training input for the DL modeling network. The *DLN* should be able analyze what are the combinations of optimization techniques that could maximize a certain criterion. The *DLN* should also infer the connection between the optimization techniques used in previous models to create new models that could even achieve better performance than existing models. However, it needs a huge number of successful optimization models to train such *DL* network.

**DSS**: The *DSS* simply retrieves existing models based on criteria submitted by the users. If more than one model retrieved. Then, the DSS let the users elucidate their preferences to further filter the retrieved models. Finally, the user decides whether to adopt one of the filtered models or to ask the DL network to predict a new model that would fit in the user requirement. The workflow of the DSS is shown in Figure 6.

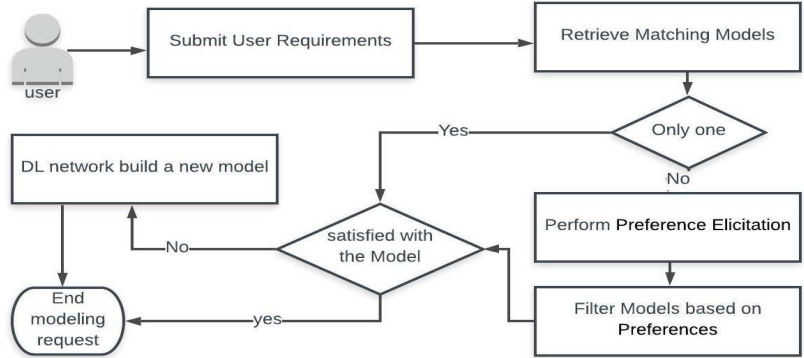

**Fig. 6:** the workflow of the DSS processes.

As shown in the figure, the selection criteria would have some mandatory requirement that would be the decision objectives. At first, relevant models are selected from the repository by querying the model that matches the user criteria. Next, the user should perform *Preference Elicitation* [36] to prioritize the desired set of objectives by giving weights to each objective in correspondence to other objectives as follows:

**Decision Criteria**: Let $Prf_j, Rel_j, Sec_j, Cst_j, Lat_j,$ and $Cmp_j$ be the values of the *performance, reliability, security, cost, latency, and complexity* respectively objectives for model $j \in J$, where the set $J$ includes all of the models in the repository. Let $w_{Prf}$ ,

$w_{Rel}$, $w_{Sec}$, $w_{Cst}$, $w_{Lat}$, $w_{Cmp}$ be the weights for the *performance, reliability, security, cost, latency, and complexity* respectively objectives. The overall score for model $j$ would be calculated as in equation 1:

$$Ovrl_j = w_{Prf} * Prf_j + w_{Rel} * Rel_j + w_{Sec} * Sec_j + w_{Cst} * Cst_j + w_{Lat} * Lat_j + w_{Cmp} * Cmp_j \tag{1}$$

The models could be ranked based on $Ovrl_j$, $j \in J$. In case that using queries didn't retrieve a model that matches the selection criteria, the DSS could send the selection criteria to the DLN modeling to create an optimization model that match the selection criteria. The DSS accordingly will select this new model as the optimum model. Then, the DSS will save a copy of the new model configurations in the repository.

## 6    Illustrative Example

In this section, we are going to explain how the framework works and what are expected inputs and outputs. A medical company ABC used the DLOM[2]. The modeling request process goes through six steps where each step represents a class in the optimization model schema presented in Table 3. At step 1, the company determines the nature of the project and related attributes as in Figures 7a. The company selected that the model cost should not exceed $14k for 10 end devices. The company has no certain specifications provided for IoT.

**Fig. 7: a)** DOLM[2] Modeling Request Step 1                **b)** DOLM[2] Modeling Request Step 2

Then, at step 2, the company selects the attributes of the *Main DLN* as shown in Figure 7b. Repeatedly, the user goes through the six classes of the *DL* modeling schema where each class represents a GUI interface. In the following steps, the user is asked if there are any preferences in specifying cloud server, end device specifications, and minimum accepted performance. Based on the requirements submitted by user, the framework builds a SPARQL query that encompasses all the requirements and turn them to query parameters to search for the matching models from the repository as shown in query 1, where "*apparea*" is the business focus and "*cost*" is the representing the budget.

**Query 1**:`SELECT *`
`WHERE {  ?apparea a type: apparea ;  ?cost a type: cost FILTER (`
`?no >= ?Nodevice)`

The query retrieved three models that match the requirements submitted by the user. So, further refinement for results should be performed. In the next step, *Preference Elicitation* is performed to identify the user top priorities. For each Objective, a pairwise comparison operation is performed to determine the weights of each of the six objectives $w_{Prf}$, $w_{Rel}$, $w_{Sec}$, $w_{Cst}$, $w_{Lat}$, $w_{Cmp}$.

Each pairwise comparison has four possible weights of *Equal Importance, Weak Importance, Stronger Importance, Absolute importance* as shown in Figure 8 a. For example: performance has a preference over the cost and complexity, while latency has a weaker importance than security and cost. Substituting the values of the six rating dimensions $Prf_j, Rel_j, Sec_j, Cst_j, Lat_j, and\ Cmp_j$ respectively to equation 1 using the weights submitted by the user.

**Fig. 8 a**: Preference Elicitation Step

**Fig 8.b:** Selected Model Results

Next, the DSS select the model with the highest $Ovrl_j$ as in Figure 8 b. However, the framework let the user decide whether to adopt the suggested model or to use the *DLN* to suggest a new model. If the user selected to build a new one, the requirements submitted by the user are sent to the *DL Modeling Network* to build a new model.

## 7    Discussion and Conclusion

Despite the complexity of *DL* models, it is no longer unreasonable to find an *IoT* device that can perform *DL* tasks. Thanks to *DL* and *IoT* optimization techniques which compress and accelerate *DL* models to fit into the tight computational tunnel of *IoT*s as end and edge devices. However, there are a myriad of optimization techniques but there is no certain criteria thar selects the best optimization techniques to be applied to an *IoT*. In addition, there are no certain optimization approach that could achieve all possible optimization objectives. This paper has two objectives. The first is to develop a DL for IoT optimization schema that joins the required optimization methods to ease model creation, selection, reuse and sharing. In addition, the paper put the abstract constituents to a *DL* optimization model management framework that could help the selection and creation of new *DL* optimization models according to user goals.

Thus, instead of using the mix and match methodologies to select the best optimization methods, the framework selects or creates new models based on the modeling history stored in the framework repository. However, there is some limitations in the paper, the first is that the proposed framework is considered an initial abstract design iteration and it needs further work to be completed. Future work is to follow an action research approach and ground theorizing to complete developing the framework. In other words, the design of the framework should be revisited to reflect the theoretical, strategic and business objectives to adopt *DLN*s for *IoT*s. Then, the instantiation and evaluation of the framework should reflect the best theoretical ground of the design.